# Accurate and Data-Efficient Micro-XRD Phase Identification Using Multi-Task Learning: Application to Hydrothermal Fluids

Yanfei Li[1†], Juejing Liu[1,2†], Xiaodong Zhao[1], Wenjun Liu[3], Tong Geng[4], Ang Li[1*], Xin Zhang[1*]

[1] Pacific Northwest National Laboratory, Richland, Washington 99354, United States

[2] Materials Science and Engineering Program, Washington State University, Pullman, Washington 99164, United States

[3] Advanced Photon Source, Argonne National Laboratory, Lemont, Illinois 60439, United States

[4] Department of Electrical and Computer Engineering, University of Rochester, New York, 14627, United States

[†] These authors contributed equally

[*] Corresponding Authors X.Z. (xin.zhang@pnnl.gov) and A.L. (ang.li@pnnl.gov)

**Abstract**

Traditional analysis of highly distorted micro-X-ray diffraction (μ-XRD) patterns from hydrothermal fluid environments is a time-consuming process, often requiring substantial data preprocessing and labeled experimental data. This study demonstrates the potential of deep learning with a multitask learning (MTL) architecture to overcome these limitations. We trained MTL models to identify phase information in μ-XRD patterns, minimizing the need for labeled experimental data and masking preprocessing steps. Notably, MTL models showed superior accuracy compared to binary classification CNNs. Additionally, introducing a tailored cross-entropy loss function improved MTL model performance. Most significantly, MTL models tuned to analyze raw and unmasked XRD patterns achieved close performance to models analyzing preprocessed data, with minimal accuracy differences. This work indicates that advanced deep learning architectures like MTL can automate arduous data handling tasks, streamline the analysis of distorted XRD patterns, and reduce the reliance on labor-intensive experimental datasets.

**Introduction**

X-ray diffraction (XRD) is a versatile technique for obtaining crystallographic information from various materials, including crystal structures, phase identification, and phase ratios. Its wide range of applications in materials science, geophysics, geochemistry, and biology underscores the importance of crystallographic data and XRD's accessibility.[1-3] Micro-XRD (μ-XRD) further enhances XRD capabilities by using a focused X-ray beam (as small as tens of microns) to collect diffraction patterns from a small area on the sample.[4,5] This enables detailed XRD mapping, facilitating studies of spatial phase distribution and interaction.[6,7] Combining μ-XRD with other techniques, such as in-situ experimentation, allows researchers to observe the formation and spatial migration of phases during chemical reactions.[8,9]

While μ-XRD is an established tool for mapping materials, especially minerals, data analysis poses significant challenges. Due to its smaller beam size, μ-XRD yields weaker diffraction signals than conventional XRD. Although high-flux synchrotron X-ray sources and improved detectors have enhanced the signal-to-noise ratio for solid samples, measurement of minerals within liquids – particularly in hydrothermal fluid conditions – remains problematic.[10-15] Here, X-ray absorption by the liquid phase, along with crystal distortions (e.g., defects, preferred orientation, poor crystallinity), also degrade the signal.[16,17] Moreover, since the beam size of the μ-XRD is approaching the size of a single crystals, the powder diffraction statistics (obtaining signals from thousands of crystals) are compromised amplifying the aforementioned adverse effects. Consequently, μ-XRD patterns collected in hydrothermal fluid environments differ markedly from those obtained via standard XRD. This discrepancy places a significant time and expertise burden on phase identification using conventional XRD patterns and crystallographic datasets.

The time-consuming and specialized expertise required for complex μ-XRD data analysis make it a prime candidate for deep neural network (DNN) techniques. DNN-based XRD analysis has proven successful in phase identification using patterns generated from crystallographic structures.[18-21] These models are trained on synthetic XRD patterns, assuming similarity to experimental data – a valid assumption for solid samples under standard conditions. However, the previously discussed distortions introduce significant discrepancies between μ-XRD patterns and their theoretical counterparts, undermining the foundation of conventional DNN-based XRD analysis. In our previous work, we addressed this by using a hybrid dataset approach, incorporating a small amount of labeled experimental data into the training set.[17] This strategy enabled successful in-situ phase identification within a hydrothermal fluid environment. Despite this success, there is a need to further reduce human intervention in dataset building, including data preprocessing, manual analysis, and labeling.

We introduce a multi-task learning (MTL) approach for μ-XRD analysis. MTL improves efficiency and accuracy by simultaneously learning multiple related tasks, especially when those tasks share underlying patterns. It also helps reduce overfitting by implicitly regularizing the model.[22, 23] Our MTL approach effectively identifies multiple phases within μ-XRD data by training a shared feature extraction network followed by task-specific branches. This allows the model to learn more generalizable representations of XRD patterns, including characteristics like peak intensity, location, and shape. Additionally, MTL's ability to extract these general features offers the potential to identify compounds not included in the training dataset. This is achieved by analyzing the similarity of feature representations between known and unknown compounds.

We tested our MTL model on experimental μ-XRD mapping datasets from a $LaCl_3$-$CaCO_3$ hydrothermal fluid system (200°C), demonstrating its efficiency, robustness, and accuracy for analyzing complex real-world samples. Notably, our MTL approach represents a novel application in DNN-based XRD analysis. This innovation contributes to the model's strong performance, achieving over 70% accuracy using synthetic data, a performance that significantly improved to 90% with the addition of a small amount of labeled experimental data. These results highlight the power of MTL to enhance μ-XRD analysis, offering a promising solution to the challenges of phase identification in dynamic environments.

**Methodology**

This section outlines our multi-task learning (MTL) approach for analyzing μ-XRD data. We applied our method to datasets from a $LaCl_3$-$CaCO_3$ hydrothermal fluid system (200 °C), focusing on identifying three phases: La-bastnaesite, calcite, and rhenium metal. Our MTL model was designed to simultaneously detect the presence of these phases within the μ-XRD patterns.

Prior to analysis, we implemented a two-step data preprocessing procedure (**Figure. 1**) to normalize and align data across different datasets. These normalized patterns were then analyzed by our multi-task deep neural network to determine the presence or absence of the target phases. Detailed descriptions of these steps are provided in the following subsections.

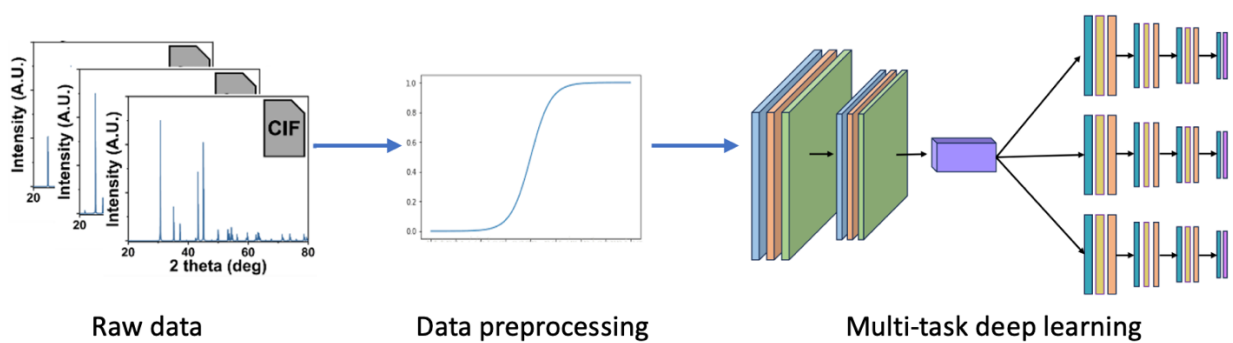

**Figure 1**. Illustration of MLT model training procedure, including synthetic XRD pattern generation, preprocessing of synthetic patterns, and training of MLT models.

**Implementation**

The code for this study was implemented in Python 3.8.17. The python packages PyTorch, NumPy and SciPy were utilized.[24-26] NumPy and SciPy were imported to preprocess the raw data, while PyTorch was used to define, train, and test the neural network.

**Datasets**

To analyze μ-XRD data collected from a hydrothermal fluid environment, three datasets are constructed: synthetic (S) dataset, experimental masked (EM) dataset and experimental unmasked (EU) dataset. For $ith$ data pair $[X_i, L_i]$, $X_i$ is the 1D μ-XRD data, $L_i$ is the label with a vector of $[b_i, c_i, r_i]$, where $b_i, c_i, r_i \in \{0,1\}$, $b_i, c_i, r_i$ are binary values, indicating whether the pattern of bastnaesite ($b_i$), calcite ($c_i$) and Re metal ($r_i$) are presence (= 1) or absence (= 0) in data $X_i$. For instance, consider data pair $[X_2, L_2]$ where $L_2 = [b_2, c_2, r_2] = [1,1,0]$. This indicates bastnaesite and calcite are present in data $X_2$, while Re metal is not.

**Synthetic dataset**

General Structure Analysis System-II (GSAS-II) is used to generate the S dataset.[27] This software enables the manipulation of three main features of XRD patterns: peak position, peak intensity, and peak profile shape, through parameter adjustments to simulate an XRD pattern. A diverse S dataset is generated by modifying combinations of these parameters.

In our S dataset, spectrum intersection range is from 3° to 60° with a 0.01° interval, resulting in a dimension of $X_i$ as 5701. The size of the S dataset is 1056: 384 samples labeled as [1,0,0], 385 samples labeled as [0,1,0] and 287 samples labeled as [0,0,1]. The example XRD patterns are shown in **Figure S1**.

**Experimental masked and unmasked dataset**

Two kinds of experimental 1D XRD datasets are prepared from the raw μ-XRD 2D images obtained in the LaCl$_3$-CaCO$_3$ hydrothermal fluid environment, including experimental marked (EM) and experimental unmarked (EU) dataset. The EU dataset was obtained by directly converting the 2D XRD image without masking and background removal. Therefore, its 1D XRD patterns contain a significant portion of background and noises. The 1D XRD patterns in the EM dataset were obtained after properly removing the background and noises in the 2D image through masking. The resulting patterns are cleaner containing more useful information than the patterns in the EU dataset (see **Figure S1**). There are 110 items in both the EM and EU dataset: 36 samples labeled as [1,1,0], 5 samples labeled as [0,1,0], 26 samples labeled as [0,0,1] and 43 samples labeled as [0,0,0]. Since the bastnaesite phase always crystallizes on the surface of

CaCO₃, the experimental XRD pattern solely containing bastnaesite (labeled as [1,0,0]) was not able to be obtained.

**Data preprocessing**

As outlined in the Datasets section, experimental XRD data and synthetic data exhibit fundamental differences. To ensure alignment across three datasets, two preprocessing steps are applied to the raw data, including x-unify and y-normalization. While the XRD pattern intersection 2-theta range of S dataset is from 3° to 60° with a 0.01° interval, the spectrum intersection of EM and EU datasets varies within the range of 4.65° to 24.8°. The 2-theta range and datapoint interval is determined by the experimental data quality and algorithms converting the 2D patterns into 1D patterns in GSAS-II.[27] To align the datasets, we unify the spectrum intersection of all data to span from 5° to 24° with a 0.01° interval. The 1D linear interpolation is adopted for the process. After unifying, the size of data $X_i$ is [1,1901]. After the X unifying, we normalized the intensity of XRD diffraction patterns. This is because the magnitude of diffraction amplitude in datasets varies significantly, which could lead to divergence during training. To address this issue, a common strategy is normalization. In this paper, we scale the unified data to [0,1], as shown in Eq.1.

$$X_i^{norm} = \frac{X_i - \min(X_i)}{\max(X_i) - \min(X_i)} \qquad (1)$$

After two steps of data preprocessing, the normalized data $X_i^{norm}$ is used as input for deep neural network.

**Network architecture**

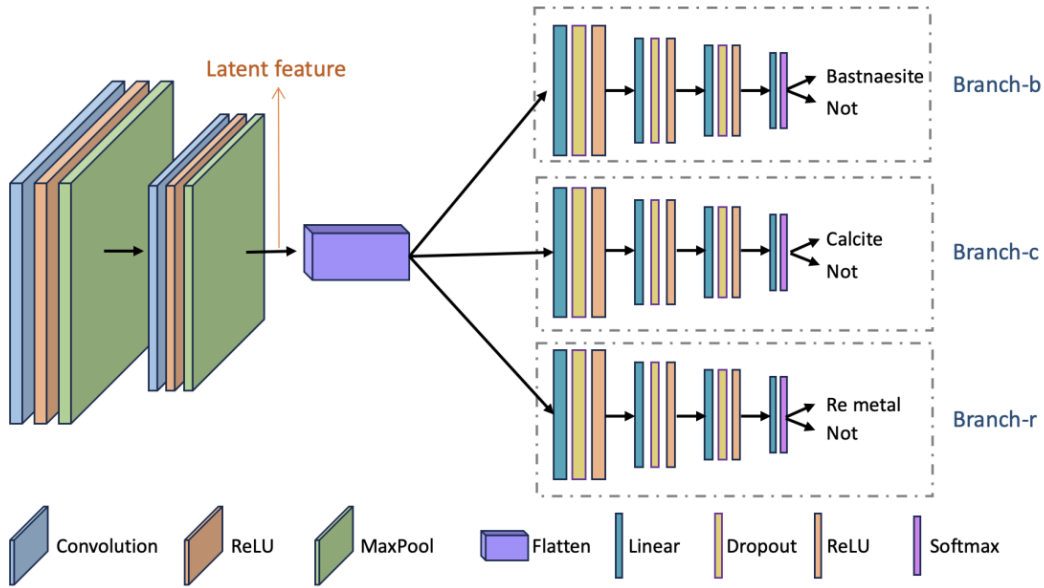

**Figure 2**. Multi-task deep learning network architecture used in this study.

We employ multi-task learning to identify the present of bastnaesite, calcite and Re metal in the XRD data. The performance of a neural network is highly dependent on its architecture. To determine the optimal network configuration, various hyper-parameters were tested, including the number of layers, kernel sizes of convolution and pooling layers, and the number of neurons in the fully connected layer, etc. After evaluating several plausible versions of neural networks, the architecture yielding the best accuracy was selected.

The model architecture, depicted in **Figure 2**, consists of two main parts. The initial part of the network is responsible for extracting general latent features, which are shared by all three tasks. The flattened latent feature representation is then directed to three branches: Branch-b, Branch-c, and Branch-r, each dedicated to a specific task. Branch-b identifies the presence of bastnaesite, Branch-c identifies calcite, and Branch-r focuses on Re metal.

The shared former part of the model comprises two pairs of 1D convolutions, each followed by a ReLU activation layer and a max-pooling layer. The 1D convolution layers serve as fundamental components for feature extraction, while the activation layers introduce non-linearities to capture complex patterns. Max-pooling layers down sample the feature maps, speeding up computation and reducing overfitting.

In our model, the kernel size of the 1D convolution is set to 11 with a stride of 2. The first convolutional layer has 8 output channels, and the second has 32. The kernel size of the first and second max-pooling is 11 and 7, respectively. The stride of max pooling is set to 2. After data preprocessing, the input data size of the model is [1,1901]. The latent features extracted by the shared part are of size [32,112], which are then flattened to [1,3584] and fed into the three branches.

Each branch follows the same architecture, consisting of four fully connected layers for prediction. The number of output neurons for these layers is set to 1024, 256, 64, and 2. To improve generalization, a dropout layer is applied after the first three fully connected layers. ReLU activation is utilized for all layers except the output layer, where softmax activation is applied to output layer. The output of the model gives probability of two classes, indicating the presence or absence of each pattern $(b, c, r)$. The class with the highest probability is considered the predicted output.

**Model training**

The model was trained end-to0-end to find the optimal parameters of the network. Cross entropy was adopted as the loss function to calculate the loss for each branch. For instance, given the input data $[X_i, L_i]$ where $L_i = [b_i, c_i, r_i]$, the output of Branch-b is $[y_0^b, y_1^b]$, indicating the absence or present of bastnaesite with probability $y_0^b$ and $y_1^b$ ( where $y_0^b + y_1^b = 1$).

As shown in Eq.2, the loss of Branch-b is computed between the data labels ($b_i$) and the output of the network. $N$ represent the total number of datasets. The loss calculations of Branch-c and Branch-r are the similar, as demonstrated in Eq.3 and Eq.4.

$$loss^b = -\sum_{i=1}^{N} \log(y_k^b) \quad k = b_i \quad (2)$$

$$loss^c = -\sum_{i=1}^{N} \log(y_k^c) \quad k = c_i \quad (3)$$

$$loss^r = -\sum_{i=1}^{N} \log(y_k^r) \quad k = r_i \quad (4)$$

The final loss is computed as the sum of the losses from three branches, as shown in Eq.5.

$$Loss = loss^b + loss^c + loss^r \quad (5)$$

The network is trained for 300 epochs on the training set, with shuffling performed before each epoch. The batch size was 60. The adaptive moment method (ADAM) with learning rate of 1e-5 and weight decay of 1e-8 was employed as the optimizer. To balance accuracy and generalization, the dropout was set as 0.2.

**Training note for experimental unmasked dataset**

As can be seen in **Figure S1**, EU dataset exhibits higher noise levels, which poses challenges for analysis. To address this and achieve better performance, three strategies were applied including weighted cross-entropy and adding noise to synthetic data.

In the EM and EU datasets, the positive samples are around 30%, resulting in class imbalances between positive and negative samples. While the impact of class imbalance on the EM dataset is minor and negligible, it cannot be ignored for the EU dataset due to its higher noise levels.

To mitigate the effects of class imbalances, weighted cross-entropy is adopted as the loss function. For instance, the loss function for Branch-b is shown in Eq.6. Compared to Eq.2, a scaling parameter $w_k$ is introduced. Through experimentation, $w_0$ is set to 1.0, while $w_1$ is set to 1.6. The loss functions for Branch-c and Branch-r are similar.

$$loss^b = -\sum_{i=1}^{N} w_k \cdot \log(y_k^b) \quad k = b_i \quad (6)$$

To further align the distribution of the S dataset with the EU dataset, we augment the S dataset by adding random noise. Gaussian noise was randomly added to each data ($X_i$) in the S dataset. The resulting noised data ($\widehat{X_i}$) served as the input to the model, as illustrated in Eq. 7. Notably, the noise was generated randomly and varied for each training epoch to enhance the diversity of the training data.

After experimentation, the mean and standard deviation of the Gaussian noise were set to 0 and 0.05, respectively. This adjustment aimed to introduce a level of randomness that would help simulate the noise present in the EU dataset, thereby narrowing the disparity between the S and EU datasets.

$$\widehat{X_i} = X_i + Gaussian(\mu = 0, \sigma = 0.05) \quad (7)$$

**Result and discussion**

Processing and analyzing µ-XRD data obtained from hydrothermal fluid environments (e.g., hydrothermal diamond-anvil cells, HDACs) presents significant challenges. The presence of a liquid phase, overexposure, preferred crystallographic orientation, and imperfect diffraction degrade data quality.[17, 28] Even within the same phase, HDAC-derived µ-XRD data deviates substantially from solid-sample patterns, often exhibiting missing peaks and lower signal-to-noise ratios (SNR). Our previous study[17] demonstrated that neural network (NN) models trained solely on synthetic XRD data were unable to accommodate these differences, preventing reliable phase identification in HDAC µ-XRD datasets. While incorporating a limited amount of labeled experimental data improved NN model accuracy, we remain interested in developing NN models capable of phase identification from experimental µ-XRD data using training based exclusively on synthetic XRD patterns. This approach would overcome the limitations imposed by the need for labeled experimental data.

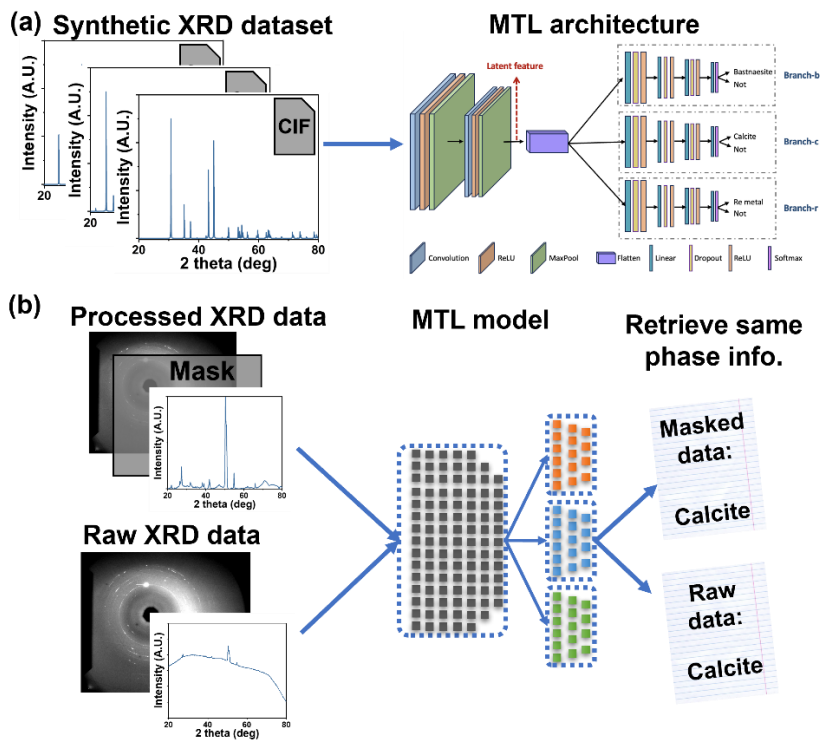

**Figure 3.** Developing multitask learning model learning from synthetic XRD pattern (a) then identifying distorted experimental pattern from micro-XRD data collected from hydrothermal fluid environment (b).

In this study, we developed a μ-XRD analysis model capable of operating without labeled experimental data (eliminating the hybrid dataset approach) and with minimal preprocessing (illustrated in **Figure 3**). We employed a comprehensive model with multi-task learning (MTL) capability offering better performance in terms of accuracy and minimize the possibility of overfitting. As outlined in **Figure 3a**, we generated a synthetic dataset containing 1056 patterns from crystallographic information files (CIF). Data augmentation expanded the dataset by mixing the three end members with different ratios (see Methodology section for details). We treated XRD patterns as input features and corresponding phases as labels for training the MTL neural network over 350 epochs. To facilitate analysis of both masked and unmasked experimental data, we developed two model variants. Unmasked data, while exhibiting lower SNR, reduces preprocessing time and effort.

We tested the real-world performance of our trained model against various XRD pattern distortions using a publicly available experimental μ-XRD mapping dataset (see **Figure 3b**). This dataset documents a $LaCl_3$-$CaCO_3$ hydrothermal fluid system at 200 °C, obtained via synchrotron radiation.[17] The system primarily exhibits three phases: La-bastnaesite, calcite, and rhenium metal. We conducted a comparative analysis to evaluate the accuracy of our experimental data-free training method against the standard hybrid dataset training approach for NN models.

Our goal was to develop a single MTL model, trained exclusively on synthetic data, to identify all potential phases within experimental μ-XRD datasets obtained from hydrothermal fluid environments. This synthetic-only approach offers several advantages. Firstly, it eliminates the need for manual analysis of experimental μ-XRD data, a resource-intensive task given the expense and time constraints of studying hydrothermal REE mineral precipitation. While manual labeling of a limited experimental dataset for inclusion in training is possible,[17] this introduces the potential for human bias. In contrast, theoretical REE mineral crystal structures are readily accessible from open source[29, 30] and proprietary databases. Synthetic XRD patterns are easily generated from these structures.[31] Therefore, training models on synthetic XRD patterns offers both time savings and bias reduction.

Training models on synthetic data for analysis of experimental data relies on a high degree of similarity between the two datasets. While viable for powder XRD (PXRD) under standard conditions,[18, 20, 21] this approach faces challenges when applied to μ-XRD studies of REE mineral precipitation in hydrothermal fluid environments. Solid PXRD samples minimize distortions caused by thermal vibration or preferred orientation, ensuring greater similarity to synthetic patterns. In contrast, the combination of solid and liquid phases in hydrothermal precipitation studies amplifies these distortions. This creates a significant divergence between experimental and synthetic XRD patterns (see **Figure S1**), hindering the performance of neural network (NN) models trained solely on synthetic data.[17] In this study, we address this gap by employing a more advanced model architecture, MTL.

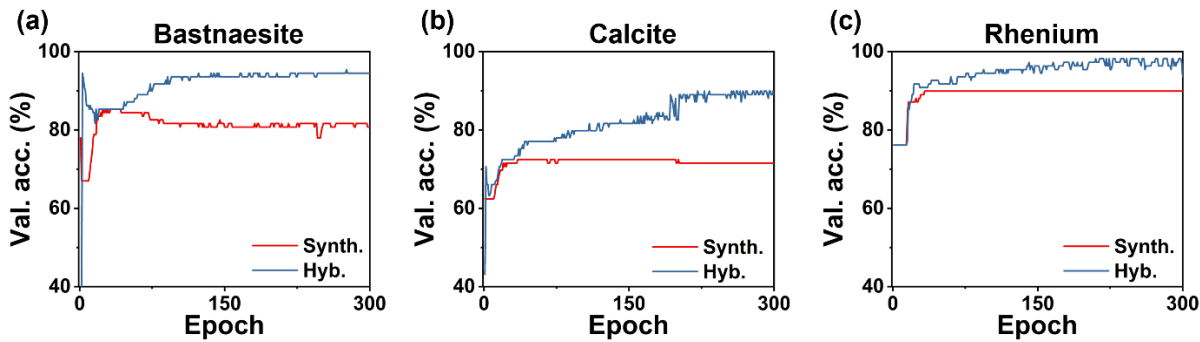

**Figure 4.** Change of test accuracy vs. epoch of training of the MTL model for analyzing masked and processed XRD data to reveal phase information of bastnaesite (a), calcite (b), and rhenium (c).

Initially, we tuned the model to analyze masked and processed µ-XRD data, focusing on the identification of bastnaesite, calcite, and rhenium metal. To mitigate the impact of low quality and overexposure artifacts common in µ-XRD (see **Figure S2**), we masked these regions during 2D to 1D data reduction. This trial aimed to explore whether a synthetic dataset, combined with the MTL architecture, could improve the accuracy of experimental µ-XRD pattern identification. We incorporated a minimal number of labeled experimental XRD patterns into the validation set. For comparison, we also prepared a hybrid dataset following our previous study,[17] with labeled experimental XRD patterns in both training and validation sets. As shown in **Figure 4a**, **b**, and **c**, the hybrid dataset model achieved high accuracies (94% bastnaesite, 91% calcite, 98% rhenium). This aligns with expectations, as the model benefits from direct exposure to labeled experimental data during training.

Intriguingly, the model trained on synthetic data demonstrates a capacity for experimental XRD pattern identification. It achieved accuracies of 81% for bastnaesite, 72% for calcite, and 91% for rhenium. While lower than the hybrid dataset model, these accuracies suffice for initial phase identification. This result suggests that despite the absence of labeled experimental XRD patterns during training, the limited inclusion of such patterns in the validation set influences hyperparameter tuning.[32]

**Table 1**. Comparison of test accuracies of DNN-based models trained by synthetic and hybrid dataset identifying phases from masked synchrotron data in previous study[17] and current study.

| Phase | Bastnaesite | Calcite | Rhenium |
|---|---|---|---|
| Synthetic data, previous study (%) | 89 | 64 | 74 |
| Hybrid data, previous study (%) | 92 | 90 | 95 |
| Accuracy difference, previous study (Δ%) | 3 | 26 | 21 |
| Synthetic data, current study (%) | 85 | 72 | 90 |
| Hybrid data, current study (%) | 95 | 92 | 98 |
| Accuracy difference, current study (Δ%) | 10 | 20 | 8 |

The advantages of the MTL architecture become evident when comparing test accuracies of models trained on synthetic and hybrid datasets (see **Table 1**). These accuracies, obtained from a test dataset excluded from training and validation, indicate generalization performance. In our previous study employing three independent CNN models,[17] accuracy discrepancies between synthetic and hybrid-trained models were significant, except for bastnaesite. Calcite exhibited a 26% difference (64% vs. 90%), and rhenium showed a 21% deviation (74% vs. 95%). Bastnaesite's smaller difference (3%, 92% vs. 89%) likely stems from its distinctive XRD pattern among the three phases.

For the models employing MTL architecture, the test accuracy differences between those trained on synthetic and hybrid datasets are notably reduced. The deviations for calcite and rhenium are 20% and 8% respectively, demonstrating a 6% and 13% improvement compared to our previous study. While the bastnaesite test difference shows a slight increase with MTL, the overall accuracies remain high at 85% (synthetic) and 95% (hybrid). We attribute the synthetic dataset model's accuracy gains to the strengths of the MTL architecture (see **Figure 2**).

A key advantage of the MTL model over traditional binary classification is its shared knowledge base, which learns common features across XRD patterns. This shared knowledge significantly improves the MTL model's accuracy, even with limited labeled experimental data. In our study, the performance difference between models trained with synthetic data and those trained with a hybrid dataset was significantly reduced with MTL (10% for bastnaesite, 20% for calcite, 8% for rhenium). This contrasts with our previous study's larger discrepancies (3% for bastnaesite, 26% for calcite, 21% for rhenium), demonstrating how the MTL architecture effectively reduces the need for extensive labeled experimental data. This is crucial, as preparing such data is time-consuming and laborious. By lowering these requirements, MTL makes ML-based XRD analysis more accessible for a wider range of applications, especially those with limited or distorted data.

Our second trial addressed two key questions: 1) Can the MTL-based model directly process unmasked, raw 1D XRD patterns obtained from 2D images, and 2) Does a weighted cross-entropy loss function enhance MTL model accuracy? Investigating the first question tackles a prevalent challenge in μ-XRD: significant background noise and distortions in 2D images.[4] These distortions, exacerbated by the solid-liquid hydrothermal environment and poor crystallinity, manifest in the 1D patterns (**Figure S1**). This typically necessitates time-consuming manual masking and background removal using various software and algorithms.[27, 33-35] Our second question addresses the dataset's inherent imbalance. Typically, 'positive' (true) examples for a given phase comprise roughly 30% of the dataset, with the remaining 70% being 'negative' (false). This can skew model accuracy when using standard cross-entropy loss, as false-positive and false-negative predictions incur different penalties. Weighted cross-entropy addresses this by equalizing penalties for both types of errors during training.[36, 37]

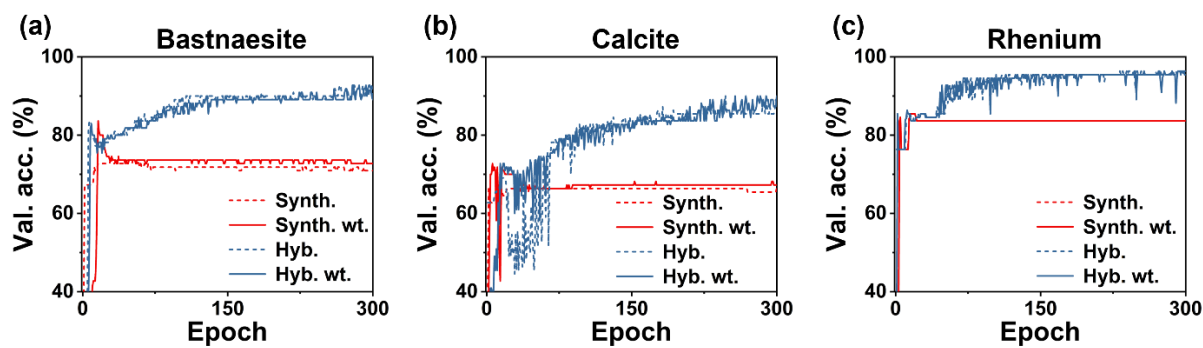

**Figure 5.** Change of test accuracy vs. epoch of training of the MTL model for analyzing unmasked and raw XRD data to reveal phase information of bastnaesite (a), calcite (b), and rhenium (c).

Our analysis reveals that utilizing unmasked, raw data during model training negatively impacts accuracies for both synthetic and hybrid datasets (see **Figure 5**). With the hybrid dataset, final accuracies decreased by 1.9% (94.5% vs. 92.6%) for bastnaesite, 2.4% (91.5% vs. 89.1%) for calcite, and 2.2% (98.2% vs. 96.0%) for rhenium when compared to models trained on masked, processed data. This finding suggests that data processing and masking might be dispensable when employing a hybrid dataset, which already includes labeled experimental data during training and validation. Furthermore, introducing weighted cross-entropy did not significantly improve accuracies for any of the three phases. This demonstrates the MTL model architecture's inherent ability to manage the imbalanced nature of the training and validation datasets.

The unmasked, raw data had a more profound negative impact on models trained with the synthetic dataset. With standard cross-entropy loss, accuracies dropped by 10.7% for bastnaesite (80.8% vs. 70.1%), 5.1% for calcite (70.5% vs. 65.4%), and 6.4% for rhenium (90.0% vs. 83.6%) compared to those trained on masked, processed data. This likely stems from the absence of labeled experimental data in the training set. Unmasked data introduces significant background noise, further exacerbating the divergence between synthetic and experimental patterns. Implementing weighted cross-entropy marginally improved accuracies by 1.8% for bastnaesite and 1.9% for calcite. Rhenium identification remained unchanged (83.6%) regardless of loss function type.

**Table 2.** Comparison of test accuracies of DNN-based models trained by synthetic and hybrid dataset with cross entropy and weighted cross entropy loss function identifying phases from unmasked synchrotron data.

| Phase | Bastnaesite | Calcite | Rhenium |
|---|---|---|---|
| Synthetic data, masked data, cross entropy (%) | 85 | 72 | 90 |
| Synthetic data, cross entropy (%) | 78 | 69 | 84 |
| Synthetic data, weighted cross entropy (%) | 84 | 72 | 85 |
| Weighted cross entropy vs. cross entropy (Δ%) | 7 | 3 | 6 |
| Best unmasked vs. masked (Δ%) | -1 | 0 | -5 |
| Hybrid data, masked data, cross entropy (%) | 95 | 92 | 98 |
| Hybrid data, cross entropy (%) | 91 | 89 | 94 |
| Hybrid data, weighted cross entropy (%) | 93 | 90 | 96 |
| Weighted cross entropy vs. cross entropy (Δ%) | 2 | 1 | 2 |
| Best unmasked vs. masked (Δ%) | -2 | -2 | -2 |

Test dataset results further demonstrate the advantages of weighted cross-entropy, particularly for models trained on the synthetic dataset. In this case, we observed test accuracy improvements of 7% for bastnaesite, 3% for calcite, and 6% for rhenium identification. Conversely, models trained with the hybrid dataset exhibited minimal gains with weighted cross-entropy: 2% for bastnaesite, 1% for calcite, and 2% for rhenium. This strongly suggests that weighted cross-entropy becomes crucial when labeled experimental data is insufficient for inclusion in both the training and validation sets.

Remarkably, MTL models demonstrate proficiency in identifying phases even within unmasked, raw XRD patterns. For models trained on the synthetic dataset, test accuracy differences between the best models analyzing unmasked vs. masked data were minimal: -1% for bastnaesite, 0% for calcite, and -5% for rhenium. The hybrid dataset exhibited a similar trend, with -2% differences across all three phases. This compelling observation implies that proper training strategies and techniques, coupled with the MTL architecture, can potentially eliminate the need for time-consuming masking and preprocessing of 2D and 1D XRD patterns.

**Conclusion**

In this study, we employed deep learning and the advanced MTL architecture to improve the analysis of highly distorted μ-XRD patterns collected in hydrothermal fluid environments via synchrotron radiation. Our goal was to minimize reliance on labeled experimental data and preprocessing (masking), which are resource-intensive tasks. We included only a minimal number of labeled experimental XRD patterns within the validation dataset. When analyzing masked and processed data, MTL models simultaneously identifying all phases outperformed three ordinary CNN-based binary classification models in models based solely on synthetic data. Additionally, accuracy discrepancies between synthetic and hybrid dataset models were reduced using the MTL architecture compared to those employing ordinary CNN architectures. We further optimized MTL models to process unmasked, raw μ-XRD patterns. Weighted cross-entropy loss significantly improved test accuracies for synthetic-trained models by 3-7%. In contrast, hybrid-trained models saw minimal gains, at most 2%. Intriguingly, MTL models tuned for unmasked, raw XRD patterns achieved accuracies close to those of models analyzing masked, processed data. The maximum accuracy differences were only 5% (synthetic dataset) and 2% (hybrid dataset). Our study demonstrates that advanced model architectures like MTL can reduce the need for labeled experimental XRD patterns during training and validation. Furthermore, masking and preprocessing of distorted μ-XRD patterns may not always be essential. This work suggests a method for streamlining XRD pattern processing and dataset creation, allowing machine learning to tackle these time-consuming tasks.

**Conflict of Interest**

The authors declare no competing financial interests.

**Acknowledgement**

The research described in this paper was conducted under the Laboratory Directed Research and Development Program at Pacific Northwest National Laboratory (PNNL), a multiprogram national laboratory operated by Battelle for the U.S. Department of Energy. Y.L., J.L., X.Z., and X.Z. acknowledge support from Advanced Research Projects Agency-Energy's (ARPA-E) Mining Innovations for Negative Emissions Resource Recovery (MINER) program with award number 0002707-1515, "Re‑Mining Red Mud Waste for $CO_2$ Capture and Storage and Critical Element Recovery (RMCCS‑CER)". A portion of the work was performed with the user proposal 51382 using the Environmental and Molecular Sciences Laboratory (EMSL), a national scientific user facility at PNNL sponsored by the DOE's Office of Biological and Environmental Research. PNNL is a multi-program national laboratory operated by Battelle Memorial Institute under contract no. DE-AC05-76RL01830 for the DOE. This research used resources of the Advanced Photo Source, a U.S. Department of Energy (DOE) Office of Science user facility operated for the DOE Office of Science by Argonne National Laboratory under Contract No. DE-AC02-06CH113577.